\documentclass[12pt,a4paper, final]{iopart}

\usepackage{iopams}  
\usepackage{graphicx}
\usepackage[breaklinks=true,colorlinks=true,linkcolor=blue,urlcolor=blue,citecolor=blue]{hyperref}

\begin{document}

\title {Gold nanoflowers as efficient hot-spots for surface enhanced Raman scattering }

\author{Arun Singh Patel }
\address{School of Computational \& Integrative Sciences, Jawaharlal Nehru University, New Delhi-110067, India}

\author{Subhavna Juneja }
\address{School of Biotechnology, Jawaharlal Nehru University, New Delhi-110067, India}

\author{Pawan K. Kanaujia}
\address{Nanophotonics Laboratory, Department of Physics, Indian Institute of Technology Delhi, New Delhi-110016, India}

	\author{G.Vijaya Prakash} 
	\address{Nanophotonics Laboratory, Department of Physics, Indian Institute of Technology Delhi, New Delhi-110016, India}
	
\author{Anirban Chakraborti}
\address{School of Computational \& Integrative Sciences, Jawaharlal Nehru University, New Delhi-110067, India}

\author{Jaydeep Bhattacharya}
\address{School of Biotechnology, Jawaharlal Nehru University, New Delhi-110067, India}
\ead{jaydpb@gmail.com}

\begin{abstract}
Gold nanoflowers are known for their use as efficient host for surface enhanced Raman scattering of organic dye molecules. In this article, gold nanoflowers  have been synthesised and  rhodamine 6 G (R6G) molecules have been used  as probe molecules.  It is found that the gold nanoflowers can detect the presence of R6G molecules upto 10$^{-10}$ M. The petals of these nanoflowers play an important role for the enhancement of Raman signals, and the intensity of Raman signals is enhanced many folds in presence of gold nanoflowers --  the enhancement factor is of the order of 10$^6$. This is explained in terms of  electromagnetic mechanism. 

\end{abstract}

\vspace{2pc}
\maketitle

\section{Introduction}

Surface enhanced Raman scattering (SERS) is a powerful analytical technique for detection of chemicals and biomolecules due to its high sensitivity \cite{kho2012,zhang2012, vo1995, lal2007, samal2013, cialla2012}. When a probe molecule is adsorbed on to the  surface of noble metal nanoparticles,  the intensity of Raman signals of the molecule is enhanced many folds. The enhancement can be of the order of 10$^{10}$ \cite{dieringer2006}. The enhancement is  attributed to either of the two possible mechanisms -- electromagnetic mechanism (EM) or chemical mechanism (CM) \cite{michaels1999,hao2004}. The electromagnetic enhancement arises due to strong electromagnetic field of metal nanoparticles associated with the localized surface plasmon resonance (LSPR) \cite{tan2013,itoh2003}. Colloidal metal nanoparticles of silver and gold are used for SERS based sensors \cite{qian2008,mulvihill2009}. These coin metals posses high density of free electrons which cause the enhancement of electromagnetic field near the surface of these nanoparticles. The enhanced field plays an important role for the enhancement of Raman signals from the probe molecules \cite{xu1999}. The chemical mechanism arises due to charge transfer between probe molecules  and host material \cite{moore2012}. The enhancement factor for EM is of the order of 10$^8$ - 10$^{12}$, whereas the enhancement factor for CM is as low as 10$^2$   - 10$^3$. The other factor which often plays a major role towards enhancement is adsorption of   probe molecules on the material surface. Noble metal nanoparticles or their thin films can be used as host for SERS signal \cite{Elias2012, fazio2013}. However, the problem with these materials is that the adsorption of some molecules on their surface is not sufficient for the enhancement to be achieved. To overcome this problem, it has been proposed that surface modification of nanomaterials be done. These modifications also have some effects on the enhanced electric field around nanoparticles due to screening effect. Thus, various different shaped nanoparticles have been synthesized and used for SERS  \cite{pustovit2006, kim2005}.

Gold nanoflowers of different shapes and sizes have been studied previously for enhancing Raman signal from different probe molecules \cite {ma2013, lu2010, qian2007}.
Niu et al.  had synthesized gold nanostars of varying number of arms and used for SERS of 4-mercaptobenzoic acid (MBA) molecules \cite{niu2015}.  Also, Xie et al. used gold nanoflowers of size 88 nm for Raman probe of Rhodamine B   molecules \cite{xie2008}. One dimensional nanostructure of gold has also been synthesized and used for different applications. Vigderman et al. have used gold nanorods and nanowires for SERS of 1,4-benzendithiol molecules \cite {vigderman2012}. 
Also, researchers have synthesized gold nanoflowers with different number of petals and used these nanoflowers as host for SERS of different molecules \cite {xu2010, jena2008, huang2012}.  In this article, we report synthesis of gold nanoflowers of  size of the order of 45 nm and these gold nanoflowers have been used as host for SERS of R6G molecules at very low concentration (100 pM).   
	\section{Experimental}
	\subsection{Chemicals and materials}
	For synthesis of gold nanoflowers, hydrogen tetrachloroaurate (III) trihydrate (HAuCl$_4$.3H$_2$O) was purchased from Sigma-Aldrich. Analytical grade L-ascorbic acid (C$_6$H$_8$O$_6$)  was obtained from Sisco Research Laboratory (SRL). Ultrapure Millipore water (18.2 M$\Omega$) was used as the solvent in all aqueous solutions and rinsing procedures. For Raman scattering study rhodamine 6 g (C$_{28}$H$_{31}$N$_2$O$_3$Cl) was procured from Sigma-Aldrich.
	\subsection{Synthesis of gold nanoflowers} 
\begin{figure}[h]
		\centering
		\includegraphics[height=4.0cm]{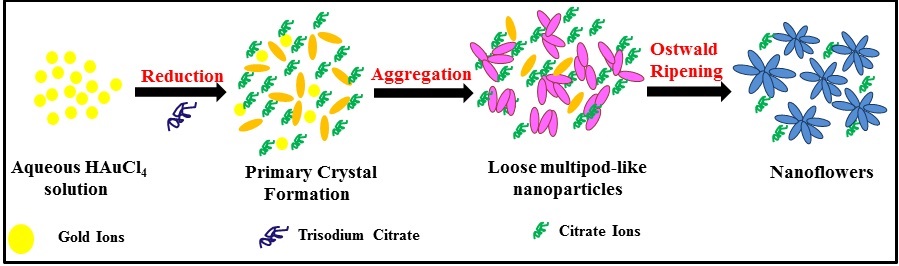}
		\caption{Schematic diagram for synthesis of gold nanoflowers.} \label{Schem}
	\end{figure}

	The gold nanoflowers were synthesized by using chemical reduction method of gold salt. In a typical synthesis procedure, 0.2 ml of 10$^{-2}$ M gold salt was mixed in 20 ml of 20 mM ice cold ascorbic acid under continuous stirring. With addition of gold salt the colorless solution of ascorbic acid is changed  to purple and finally faint blue. The mixture was allowed to stir for additional 10 min. The schematic for synthesis of gold nanoflowers is shown in figure \ref{Schem}. 
	The aqueous solution of gold salt contains gold ions which are  reduced in the presence of citrate ions, and causes nucleation for gold nanoflowers.
	\subsection{Characterization techniques}
	The optical properties of the gold nanoflowers were measured by absorption spectrophotometer in visible and near infra red region. The shape  and size of these nanoflowers were determined by transmission electron microscope  (model
	JEOL-2100F, Japan), operating at 200 kV.  The Raman spectra of rhodamine 6 g molecule were collected by Renishaw inVia
	confocal Raman spectrometer using excitation laser
	source of 532 nm and the associated grating is 2400 lines per mm. The Raman
	spectra were collected by using 50X objective lens having
	numerical aperture 0.25.
	\subsection{ Surface-enhanced Raman scattering of gold nanoflowers} 
	For surface enhanced Raman scattering of R6G molecules due to gold nanoflowers, 0.01 ml of 10$^{-9}$ M and 10$^{-10}$ M R6G solutions were mixed with 0.01 ml of gold nanoflowers solution. These mixed solutions were put on silicon substrates and dried at room temperature. 
	For comparative studies on the enhancement factor, 10$^{-3}$ M R6G solution was put on the silicon substrate.  
	
\section{Results and discussions}
\subsection{UV-Vis Absorption }
\begin{figure}[h]
	\centering
	\includegraphics[height=8.0cm]{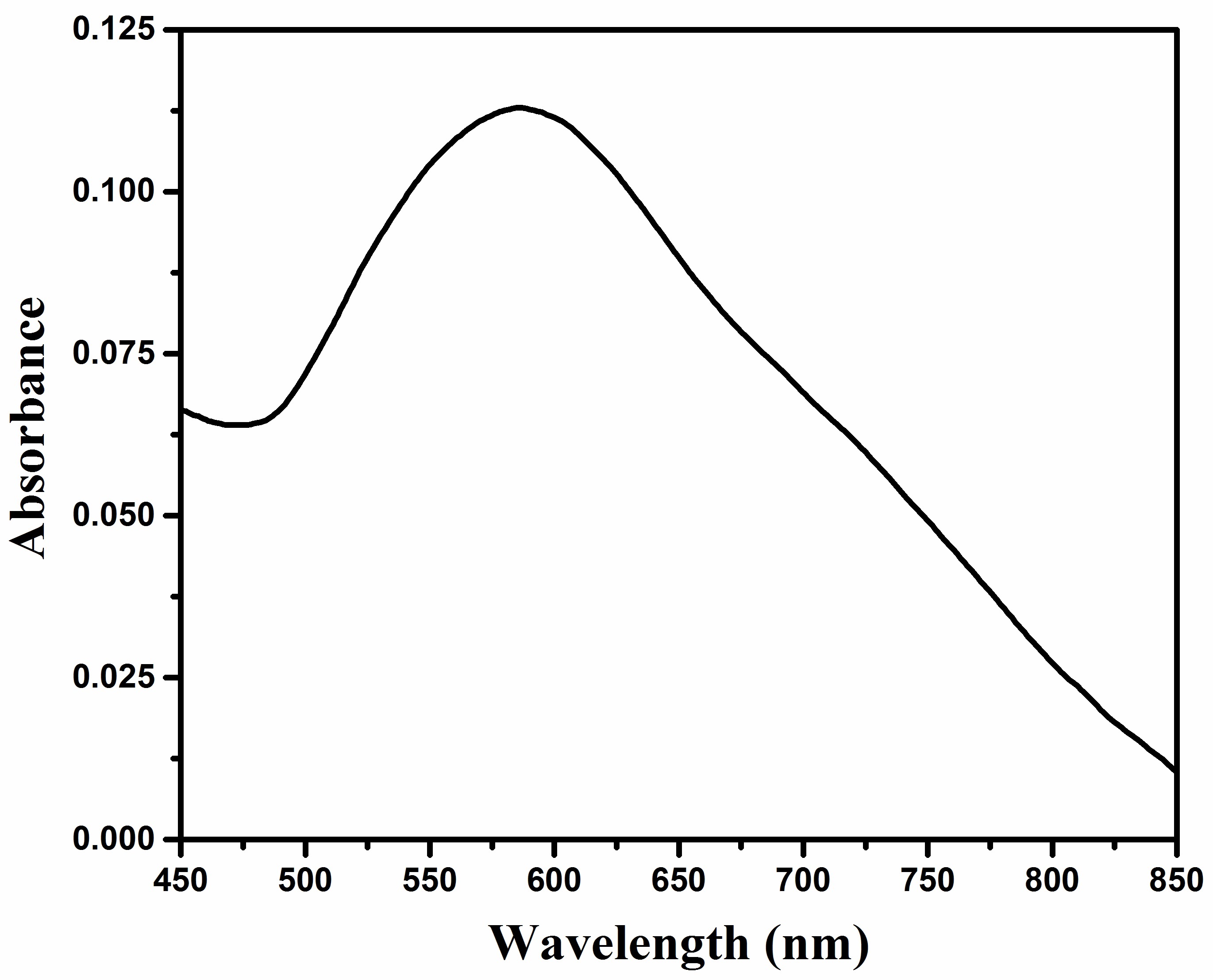}
	\caption{Absorption spectrum of gold nanoflowers.} \label{Abs}
\end{figure}

The absorption spectrum of gold nanoflowers is shown in Figure \ref{Abs}. Gold nanoflowers show two distinct absorption peaks. The peak at 550 nm is due to core of nanoflower and a hump around 700 nm originates due to free electrons which are confined in petals. These absorption peaks are known as surface plasmon resonance (SPR) peaks. The free electrons near the surface of gold nanoflowers are known as surface plasmons and when a light of particular wavelength is incident on the surface a resonance phenomenon takes place and absorption of light is observed. The oscillation frequency associated with these plasmons depends on various factor like, size, shape      of metal nanoparticles and also on the dielectric constant of the surrounding medium \cite {patel2013}. In present case, the gold nanoflowers have two distinct frequency of surface plasmons, one arising due to the core and the other due to petals.

\subsection{Transmission electron microscopy analysis}
\begin{figure}[h]
	\centering
	\includegraphics[height=12cm]{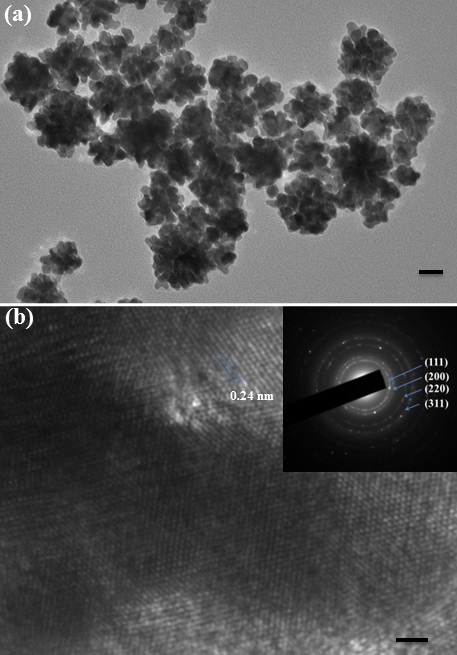}
	\caption{TEM images of gold nanoflowers (a), lattice spacing in gold (b) and inset of (b) shows the SAED pattern of gold nanoflowers; scale equals to 20 nm in (a) and 1 nm in (b).} \label{TEMNFs}
\end{figure}

The TEM images of gold nanoflowers are shown in Figure \ref{TEMNFs}.
The size of nanoflowers are of the order of 45 nm, the core size is around 35 nm, the petals are of lengths around 5 nm. Figure \ref{TEMNFs}(b) shows the lattice spacing of the gold crystal and the spacing is of the order of 0.24 nm, which corresponds to (111) plane \cite {donasong2015}. The inset of Figure \ref{TEMNFs}(b) shows the selective area electron diffraction (SAED) pattern from different lattice planes of gold. The innermost diffraction ring is associated with (111) plane; similarly other rings represents (200), (220), and (311) planes.     
\subsection{Raman study}
\begin{figure}[h]
	\centering
	\includegraphics[height=16.0cm]{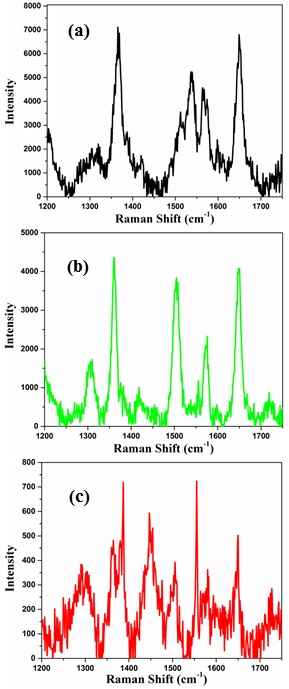}
	\caption{(a) Raman spectra of 10$^{-3}$ M R6G molecules in absence of gold nanoflowers, (b) and (c) show Raman spectra of R6G molecules with 10$^{-9}$ M and 10$^{-10}$ M concentration respectively; in presence of gold nanoflowers. }\label{Raman}
\end{figure}

The Raman spectra of R6G molecules in absence and presence of gold nanoparticles are shown in Figure \ref{Raman}. 
Figure \ref{Raman}(a) shows the Raman spectrum of R6G molecules with concentration of 1 mM, when no gold nanoflowers are there, and  the Figure \ref{Raman}(b) and (c)  represent the Raman spectra in presence of gold nanoflowers with concentrations of R6G molecules as low as 1x10$^{-9}$ M and 1x10$^{-10}$ M respectively. The intensity of Raman peak for reference sample is 7000 counts, in presence of gold nanoflowers; for 10$^{-9}$ M and 10$^{-10}$ M concentrations, the intensities are 4500 and 500 counts respectively.   These spectra confirm the Raman signal of R6G molecules can be enhanced even at very low concentration of 100 pM. The ability of gold nanoflowers on the SERS signal is calculated in terms of enhancement factor ($EF$) \cite{tao2003}
\begin{equation}
EF=\frac{I_{SERS}/C_{SERS}}{I_{N}/C_N},
\end{equation} 
where $I_{SERS}$ and $I_N$ are the intensities of a vibrational mode of R6G molecules in Raman signals, in the presence and absence of gold nanoflowers, respectively; $C$ being the corresponding concentrations of R6G molecules.   The value of $EF$ for gold nanoflowers was found to be 10$^6$. This large $EF$ is due to sharp edges of these nanoflowers. At the petals, the edge causes enhancement of electromegnetic field and this  field further interact with the adsorbed molecules and hence causes the enhancement of the Raman signal. 

The intensity of Raman signal is enhanced near the gold nanoflowers. The gold nanoflowers has two kinds of surface plasmon resonance peaks; one arising due to core of NFs and the other due to the petals. When a laser beam of 532 nm is used, at this wavelength the plasmons are  in resonance condition and hence results in strong  electromagnetic field around the Au NFs. The enhanced electromagneti field helps in  strengthening the vibrational modes of the probe molecules. All vibrational modes of R6G molecules are enhanced in the Raman peaks.    
	\section{Conclusions}
	In summary, gold nanoflowers of size 45 nm were synthesized and used as efficient hot-spots for enhancing Raman signal of R6G molecules. The Raman signals of R6G molecules in presence of gold nanoflowers were found to be enhanced up to 10$^6$ fold. The gold nanoflowers showed strong enhanced electromagnetic field surrounding the particles, which helped in the enhancement of Raman signals. The enhancement was due to electromagnetic mechanism. The gold nanoflowers could be used as SERS-based sensing of foreign molecules, up to picomolar concentration. 
	\\
	\\
	$\bf{Acknowledgement}$\\ \\	
	AC acknowledges the financial support by institutional research funding IUT (IUT39-1) of the Estonian Ministry of Education and Research. AC and ASP acknowledge financial support from grant number BT/BI/03/004/2003(C) of Government of India, Ministry of Science and Technology, Department of Biotechnology, Bioinformatics division. JB would like to thank the internal support of JNU in terms of the UPE-II grant. 
	Authors thank the AIRF, JNU for TEM  characterization and the FIST (DST, Govt. of India) UFO scheme of IIT Delhi for Raman/PL facility.

\section*{References}
\bibliography{main}
\bibliographystyle{iopart-num}

\end{document}